\documentclass{moriond}

\usepackage{caption}
\usepackage{subcaption}

\def\Journal#1#2#3#4{{#1} {\bf #2}, #3 (#4)}

\def\PRL{\em Phys. Rev. Lett.}
\def\PRD{{\em Phys. Rev.} D}

\def\CQG{\em Class. Quantum Grav.}

\def\be{\begin{equation}}
	\def\ee{\end{equation}}
\def\bea{\begin{eqnarray}}
	\def\eea{\end{eqnarray}}

\newcommand{\Photo}{\includegraphics[height=35mm]{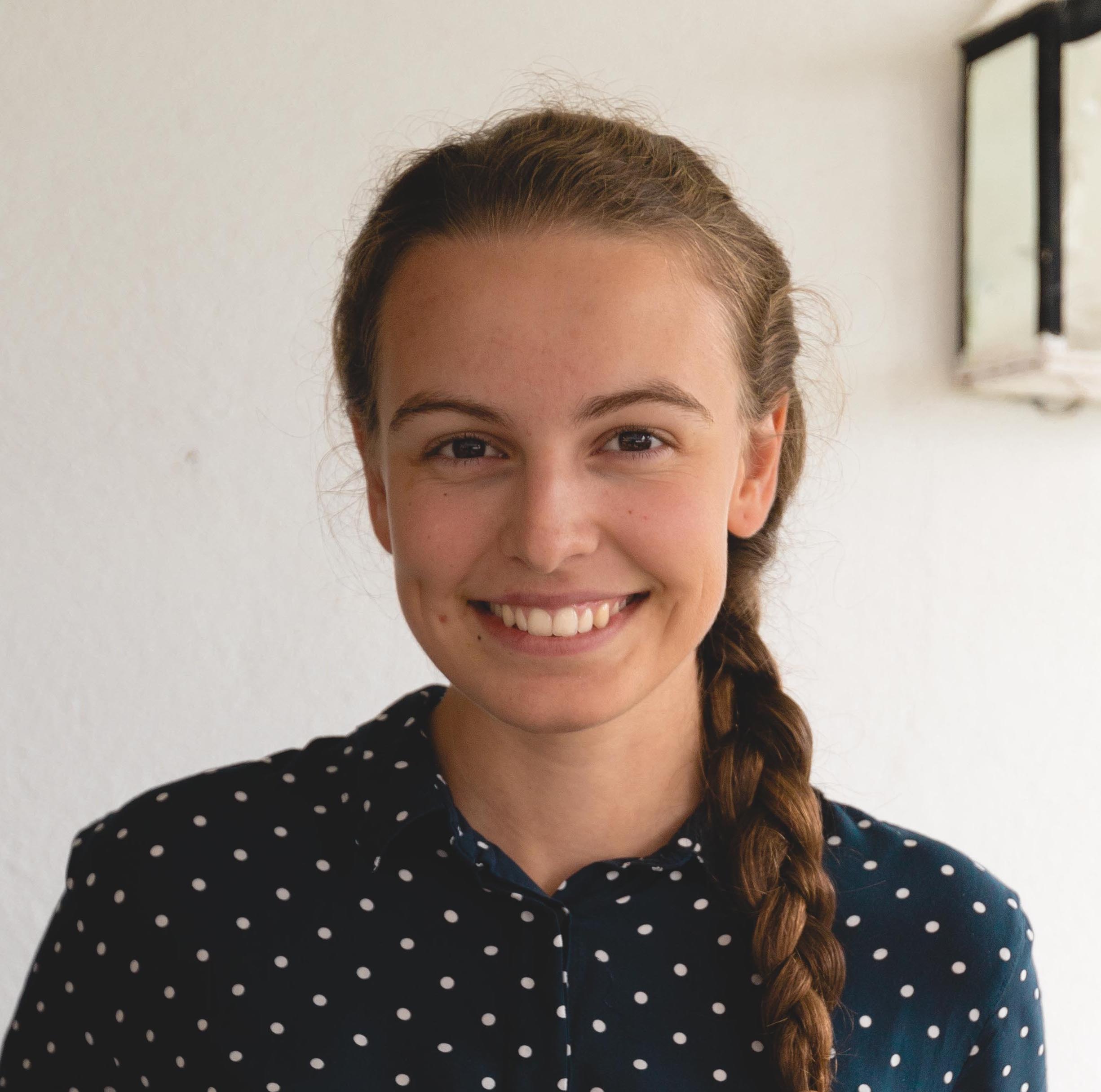}}

\begin{document}
	\vspace*{4cm}
	\title{BLACK-HOLE KICKS: A TOOL TO MEASURE THE ACCURACY OF GRAVITATIONAL-WAVE MODELS}

	\author{ANGELA BORCHERS and FRANK OHME}
	
	\address{Max Planck Institute for Gravitational Physics (Albert Einstein Institute), 30167 Hannover, Germany\\
		Leibniz Universität Hannover, 30167 Hannover, Germany}
	
	\maketitle\abstracts{
		Asymmetric binary systems radiate linear momentum through gravitational waves, leading to the recoil of the merger remnant. Black-hole kicks have attracted much attention because of their astrophysical implications. However, little information can be extracted from the observations made by LIGO and Virgo so far. In this work, we discuss how the gravitational recoil, an effect that is encoded in the gravitational signal, can be used to test the accuracy of waveform models. Gravitational-wave models of merging binary systems have become fundamental to detect potential signals and infer the parameters of observed sources. But, as the interferometers' sensitivity is enhanced in current and future detectors, gravitational waveform models will have to be further improved. We find that the kick is highly sensitive to waveform inaccuracies and can therefore be a useful diagnostic test. Furthermore, we observe that current higher-mode waveform models are not consistent in their kick predictions. For this reason, we discuss whether measuring and improving waveform accuracy can, in turn, allow us to extract meaningful information about the kick in future observations. 
	}
	
	\section{Introduction}
	Gravitational waves (GWs) emitted by asymmetric black-hole binaries carry energy, angular and linear momentum away from the system. From momentum conservation, the center of mass (CM) of the system will move in the opposite direction to the radiated momentum flux. Since the flux radiation rotates with the orbital movement of the binary, the CM will spiral with the orbital motion. In the early inspiral phase, the momentum flux is radiated isotropically on average over each orbit. As the two bodies come closer, during the last few orbits before the merger, the flux increases drastically and GWs are radiated in a preferred direction. The emission of the radiation is then abruptly cut off and drops exponentially as the remnant black hole ringdowns into a stationary state. The final black hole is thus kicked by the emission of gravitational radiation.
	\begin{figure}[tb]
		\begin{subfigure}{.5\textwidth}
			\centering
			\includegraphics[scale=0.25]{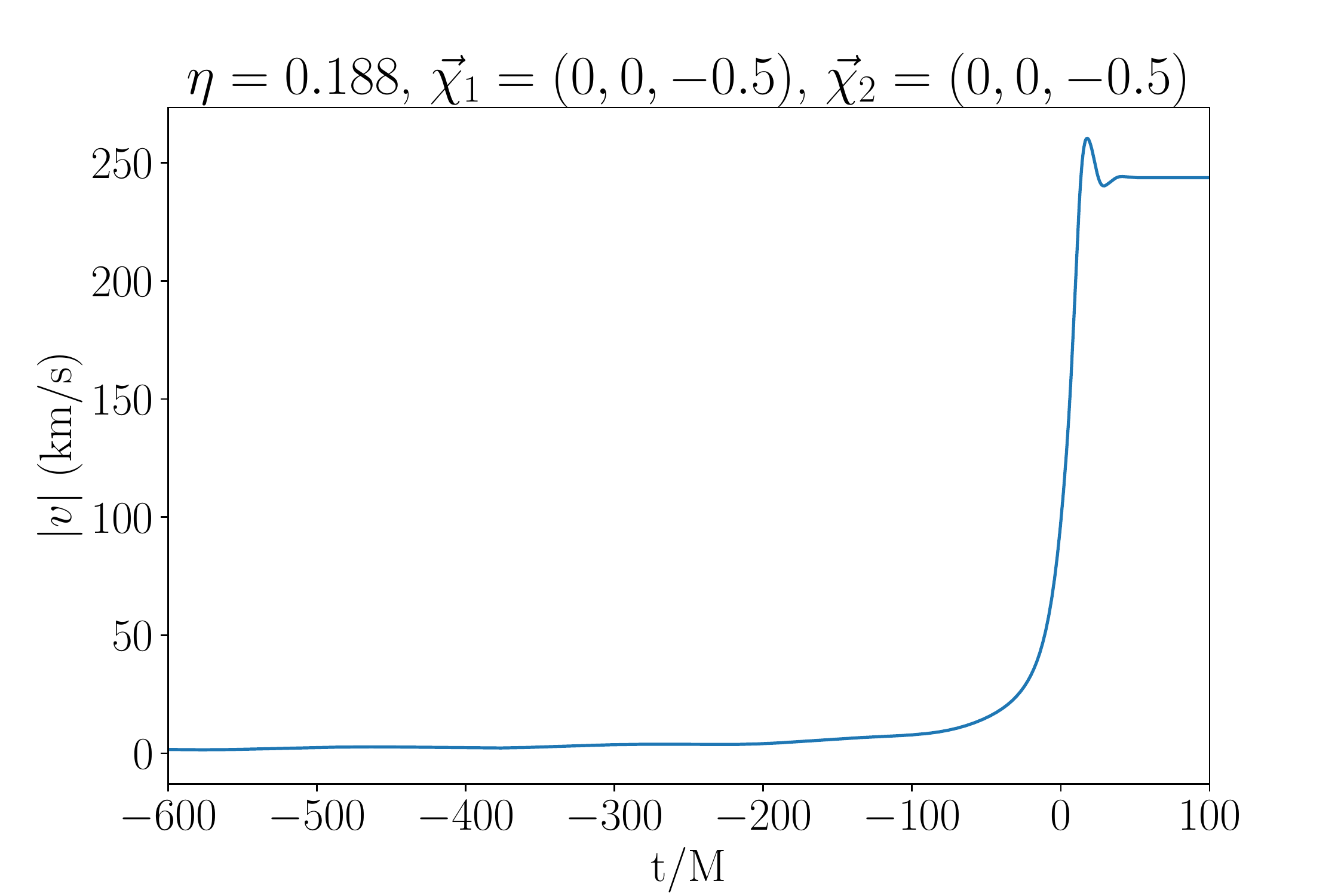}  
		\end{subfigure}
		\begin{subfigure}{.5\textwidth}
			\centering
			\includegraphics[scale=0.25]{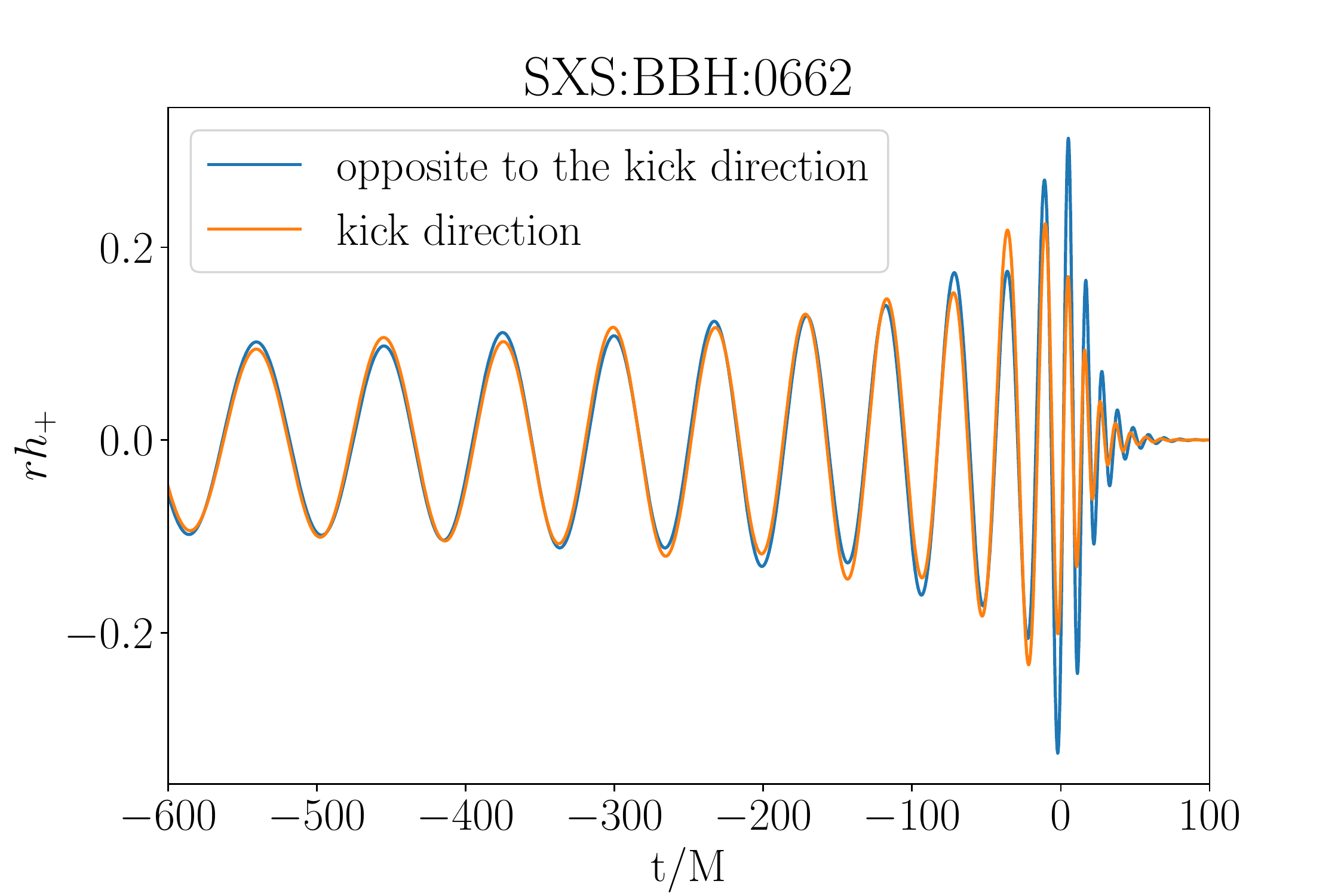}  			
		\end{subfigure}
		\caption{The left plot serves as an example of how the velocity of the CM of the binary evolves over time. The figure on the right displays the waveform asymmetries that cause the recoil for an NR precessing waveform with a kick velocity $\vec{v} = (632.587, 88.320, -3055.229)$ km/s.}
		\label{intro}
	\end{figure}
The asymmetries of the system that are responsible for the kick are imprinted in the gravitational signal. Fig. \ref{intro} shows an example for a Numerical Relativity (NR) waveform \cite{SXS}, which corresponds to a black-hole binary with a kick velocity of $\vec{v} = (632.587, 88.320, -3055.229)$ km/s, expressed in a source-centered coordinate frame, with the $z$-axis along the direction of the initial orbital angular momentum.

	\section{Morphology of the kick}
	The momentum acquired by the remnant black hole is equal to the negative of the 3-momentum carried by the radiated GWs, and is given by 
	\begin{equation}
		P_{i} = - \lim_{r\rightarrow\infty} \frac{r^{2} c^{2}}{16\pi G} \int_{-\infty}^{\infty} dt\ \oint d\Omega\ \hat{x}_{i}(\theta,\phi)\ | \dot{h}|^{2}.
		\label{momentum_balancelaw}
	\end{equation}
	It is interesting to mention that the momentum is entirely determined by the waveform, since the asymmetries that lead to the kick are encoded in the signal. In a non-precessing binary, when the spin components of the two objects are parallel to the orbital angular momentum, the kick takes place in the orbital plane. Although precessing binaries get larger kick values, here, we focus on the higher-mode description of the kick. The gravitational radiation is usually expanded in a basis of spin-weighted spherical harmonics (SWSH), 
	\begin{equation}
		h := h_{+} - ih_{\times} =  \lim_{r\rightarrow\infty} \frac{1}{r}\ \sum_{l=2}^{\infty}\sum_{m=-l}^{l} h_{l,m}(t,{\lambda}) \ _{-2}Y_{l,m}(\theta,\phi).
	\end{equation}
	Here, $h_{l,m}(t, {\lambda})$ are the radiation multipoles, which express the dependence on time and the intrinsic properties of the source, ${\lambda}$. The dependence on the orientation of the binary is described by the SWSH basis functions $_{-2}Y_{l,m}(\theta,\phi)$, where $(\theta,\phi)$ are the usual spherical angles defined in a source-centered coordinate frame. The momentum of the CM can then be expressed as a combination of the two planar coordinates,
	\begin{equation}
		{P}_{\perp} := P_{x} + iP_{y} = - \frac{ c^{2}}{8\pi G} \int_{-\infty}^{\infty} dt\ \sum_{l,m}
		\dot{h}_{l,m}(a_{l,m}\dot{h}^{*}_{l,m+1} + b_{l,-m}\dot{h}^{*}_{l-1,m+1} - b_{l+1,m+1}\dot{h}^{*}_{l+1,m+1}),
		\label{Pperp}
	\end{equation}
	where the coefficients $a_{l,m}$ and $b_{l,m}$ read
	\begin{equation}
	a_{l,m} = \frac{\sqrt{(l-m)(l+m+1)}}{l(l+1)} \quad;\quad b_{l,m} = \frac{1}{2l}\sqrt{\frac{(l-2)(l+2)(l+m)(l+m-1)}{(2l-1)(2l+1)}}.
	\end{equation}
	The kick velocity will be given by ${v}_{f} = {P}/M_{f}$, where $M_{f}$ is the mass of the remnant black hole. 
	
The momentum of the remnant black hole is a quantity that appears on an infinite set of balance laws that are predicted by full, non-linear General Relativity \cite{CBC_constraints}. Together they form an infinite set of constraints that can be applied to gravitational models of compact binaries, including NR waveforms. Yet, the momentum flux integral itself has been used to test the accuracy of the higher harmonics in the development of the latest binary black hole phenomenological models \cite{PhenomXHM}. Here, we reproduce and extend the idea of using the kick as a diagnostic tool. We find that studying the kick dependencies with respect to the intrinsic parameters of the binary is not only relevant to better understand the morphology of the kick. It is also interesting to find waveform inaccuracies over the parameter space. As an example, we show the dependency on the symmetric mass ratio $\eta = m_{1}m_{2}/(m_{1}+ m_{2})^{2}$, for two aligned-spin, higher-mode models, IMRPhenomHM \cite{PhenomHM} and IMRPhenomXHM \cite{PhenomXHM}, two models that are currently used in GW analyses. We compare their predictions to those coming from a set of NR waveforms \cite{SXS} and an NR surrogate fit \cite{surfinBH}.
	\begin{figure}[htb]
		\begin{subfigure}{.5\textwidth}
			\centering
			\includegraphics[scale=0.3]{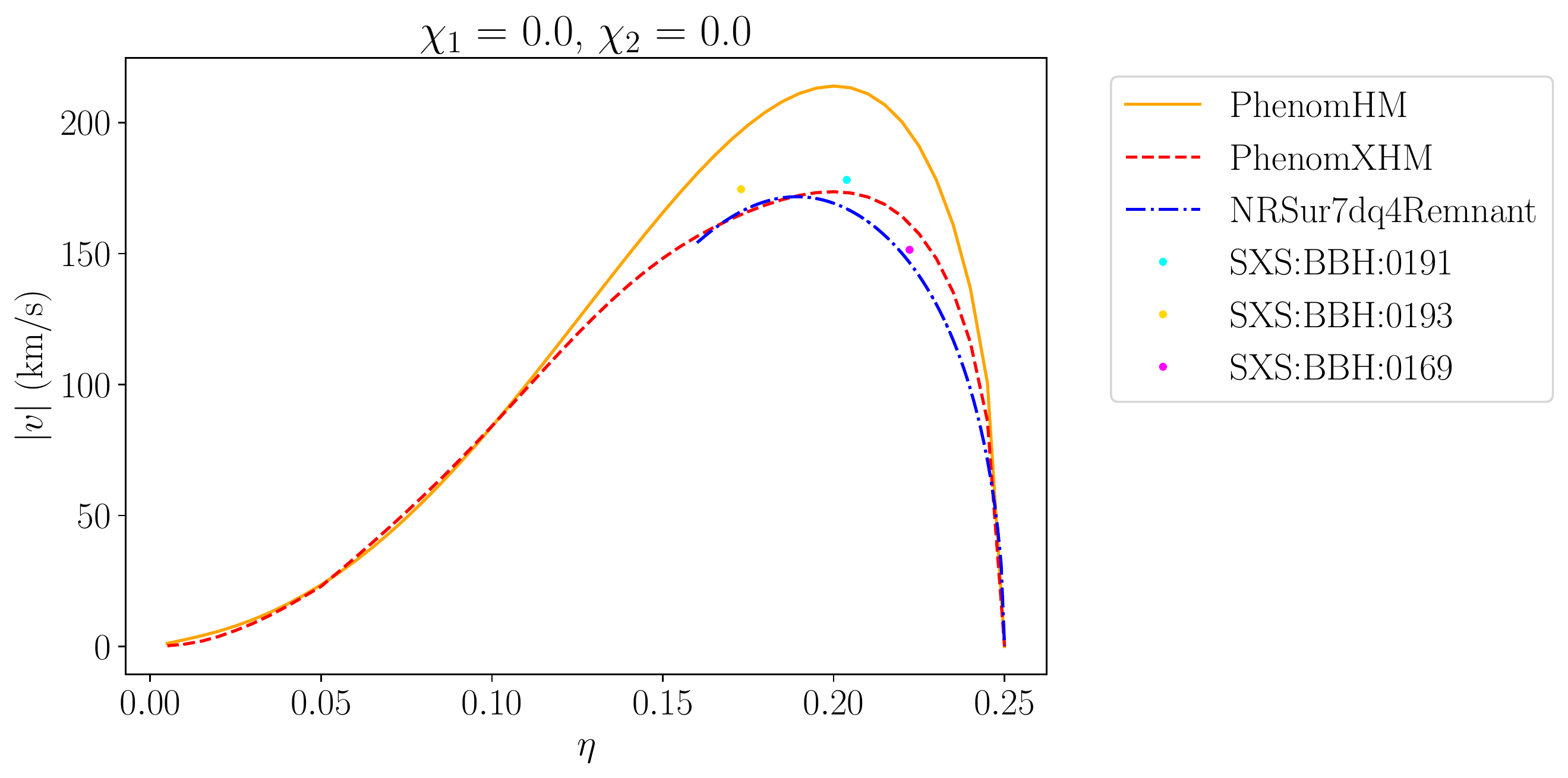}  
		\end{subfigure}
		\begin{subfigure}{.5\textwidth}
			\centering
			\includegraphics[scale=0.3]{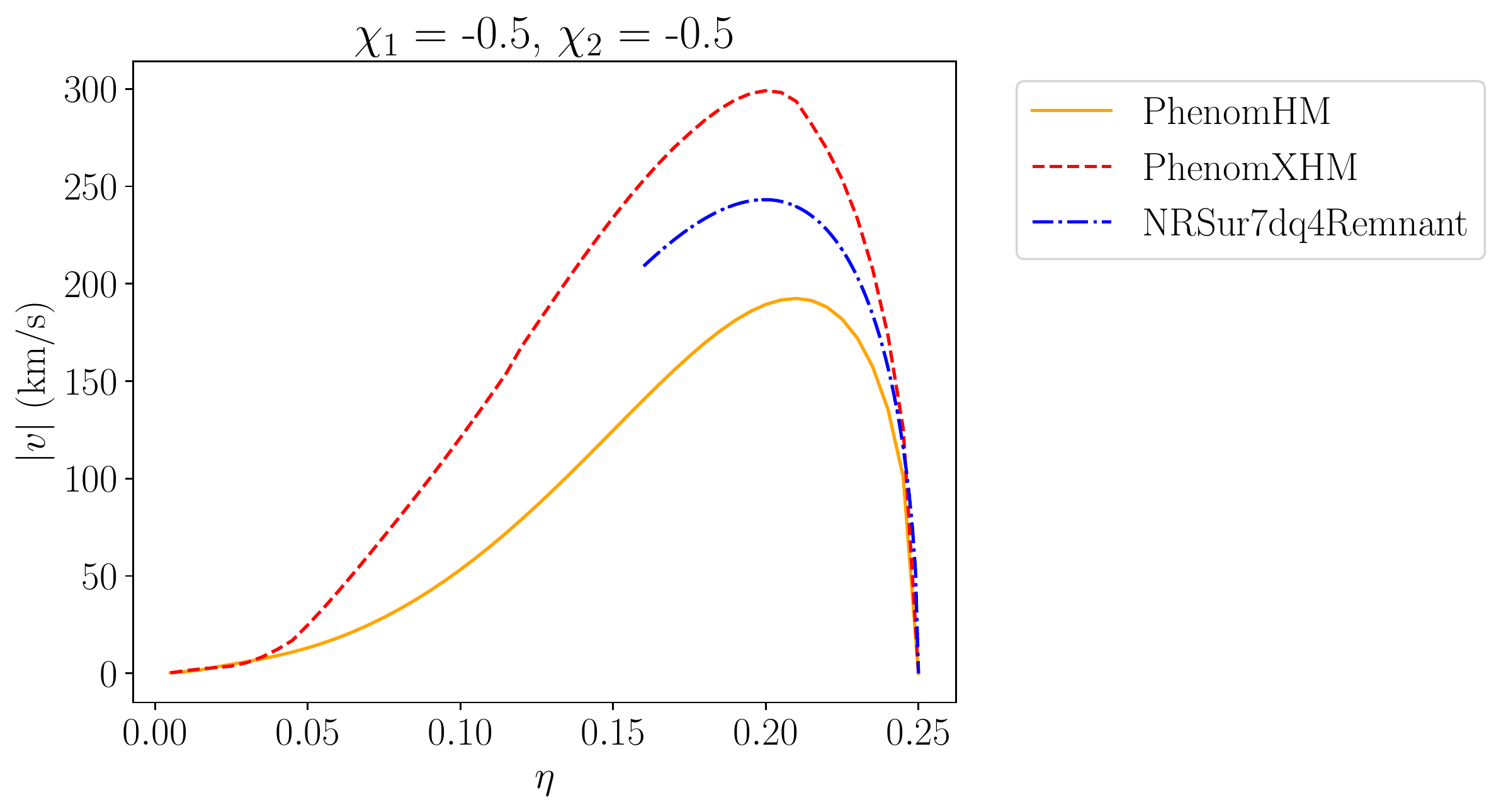}  			
		\end{subfigure}
	\begin{subfigure}{.3\textwidth}
		\centering
		\includegraphics[scale=0.25]{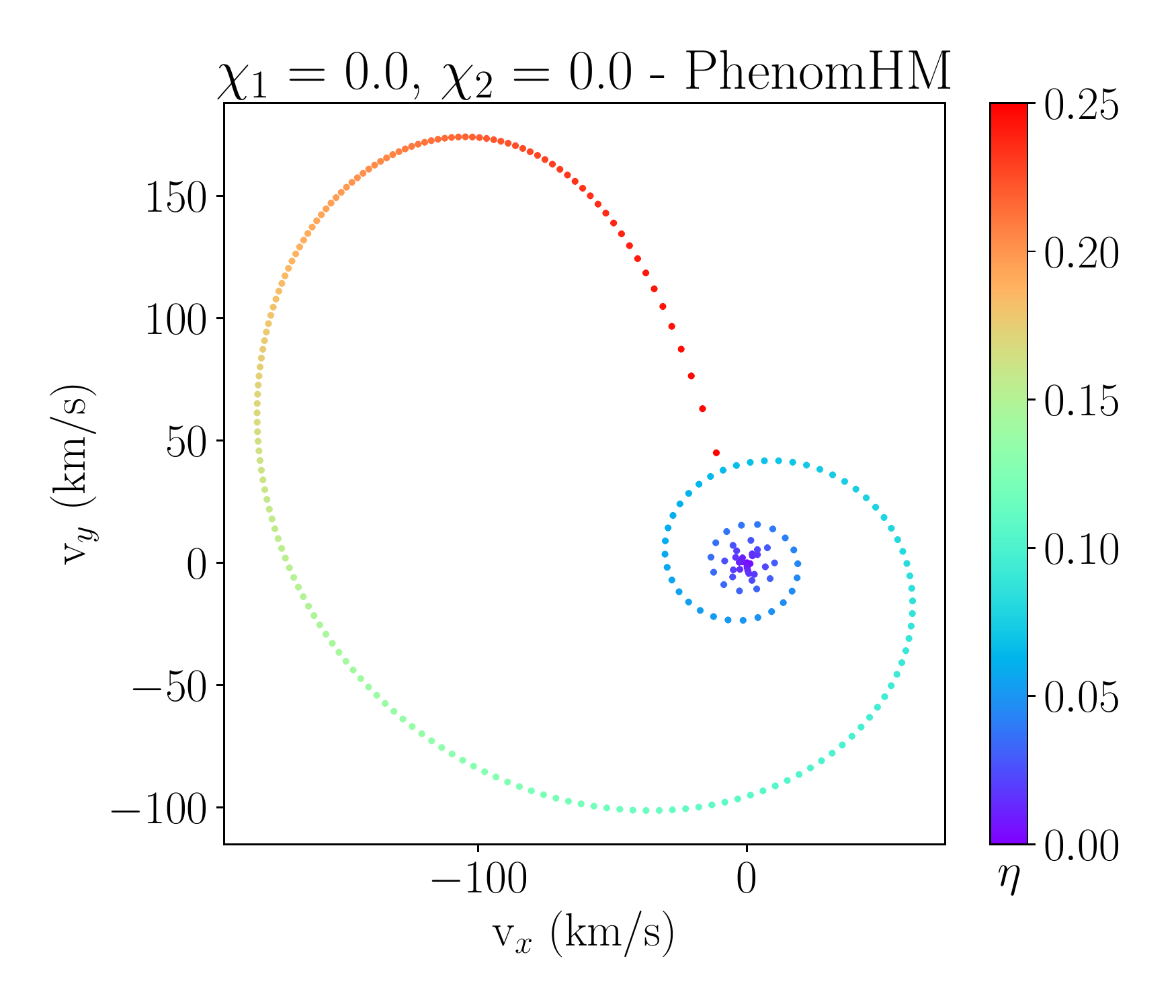}  
	\end{subfigure}
	\begin{subfigure}{.3\textwidth}
		\centering
		\includegraphics[scale=0.25]{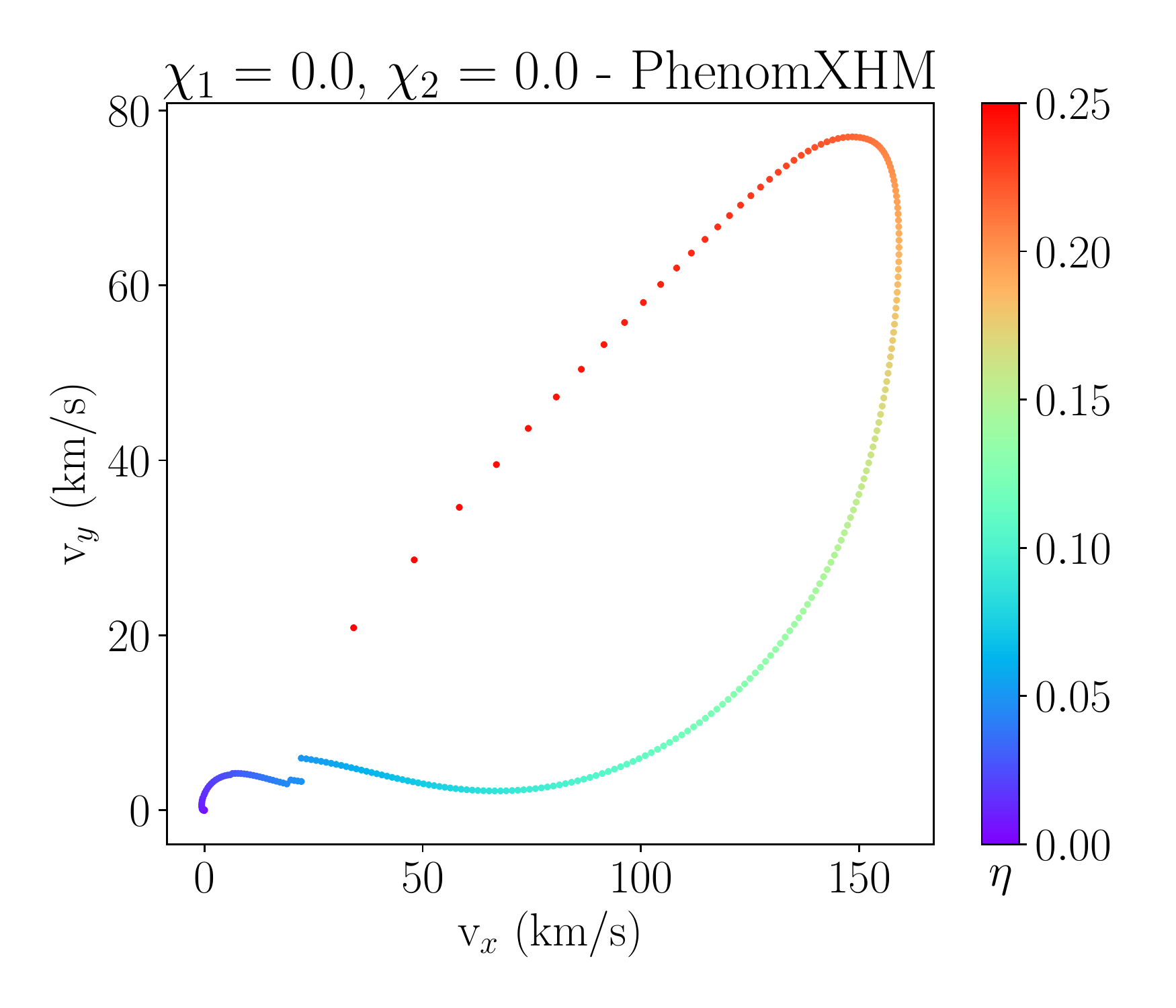}  			
	\end{subfigure}
	\begin{subfigure}{.3\textwidth}
	\centering
	\includegraphics[scale=0.25]{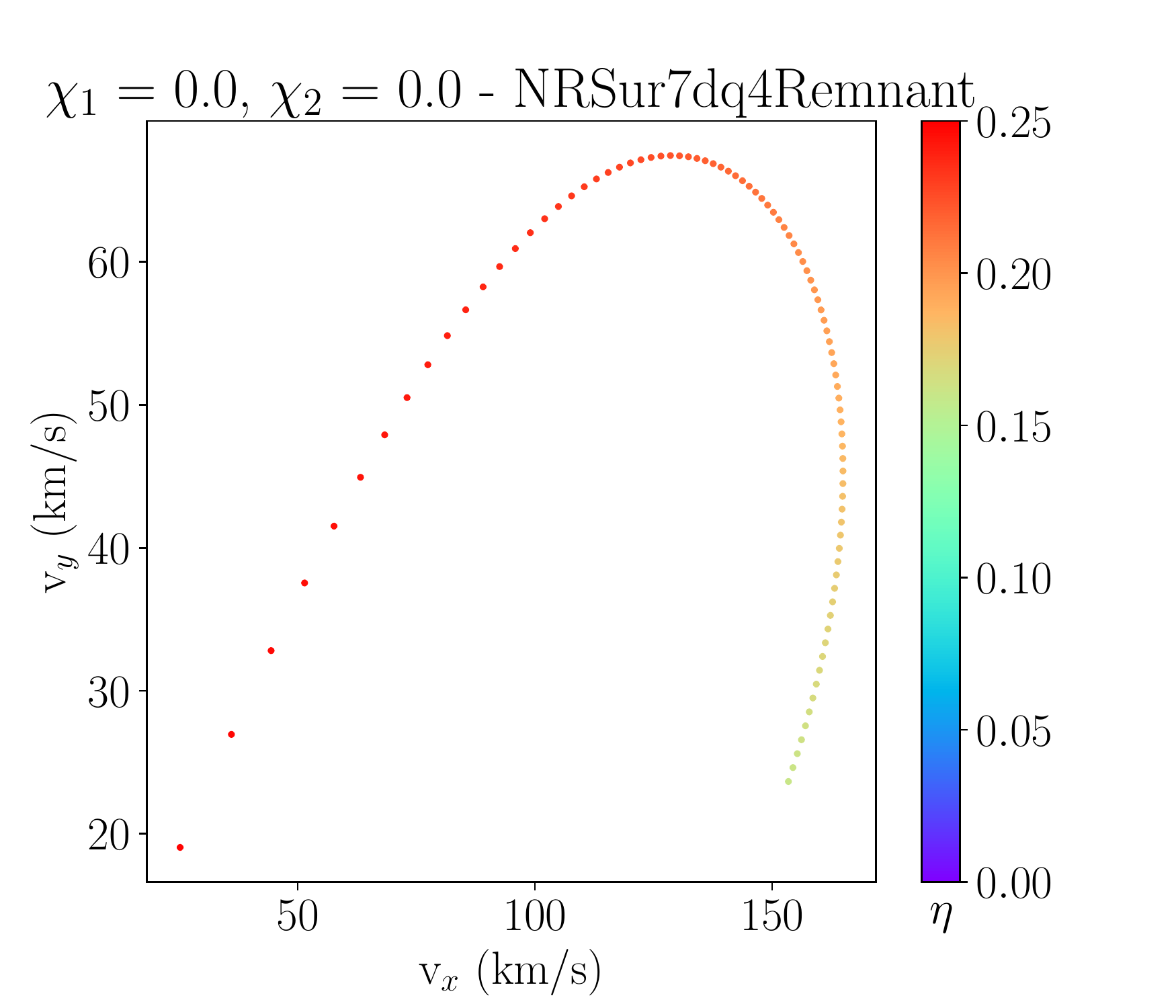}  			
	\end{subfigure}
	\caption{Symmetric mass ratio dependency of the magnitude and orientation of the kick velocity.}
	\label{eta_dependency_kick}
\end{figure}

\noindent As Fig. \ref{eta_dependency_kick} shows, the predictions of the higher-mode models differ, both, in the kick magnitude and orientation, as a function of the symmetric mass ratio. We should mention that the final direction is subject to the orbital reference phase. There is some ambiguity on how this quantity is defined for each model, so one should be careful when comparing final kick orientations. 

\section{Harmonic spectroscopy}
The recoil is actually built up by the contributions of a set of pairs of $h_{l,m}$ modes, those combinations specified precisely by Eq. \ref{Pperp}.  
\begin{figure}[b]
		\begin{subfigure}{.3\textwidth}
			\centering
			\includegraphics[scale=0.26]{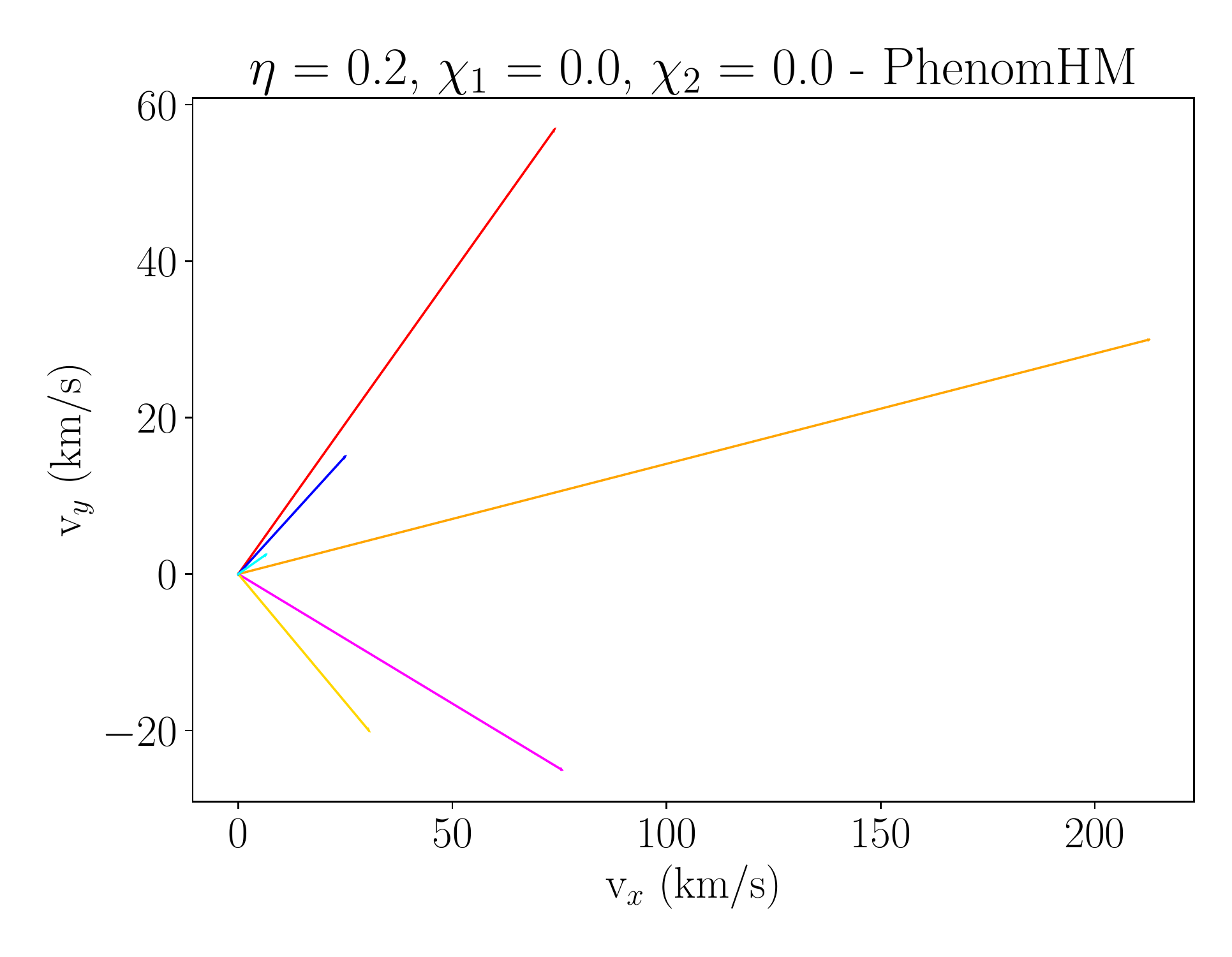}  
		\end{subfigure}
		\begin{subfigure}{.3\textwidth}
			\centering
			\includegraphics[scale=0.26]{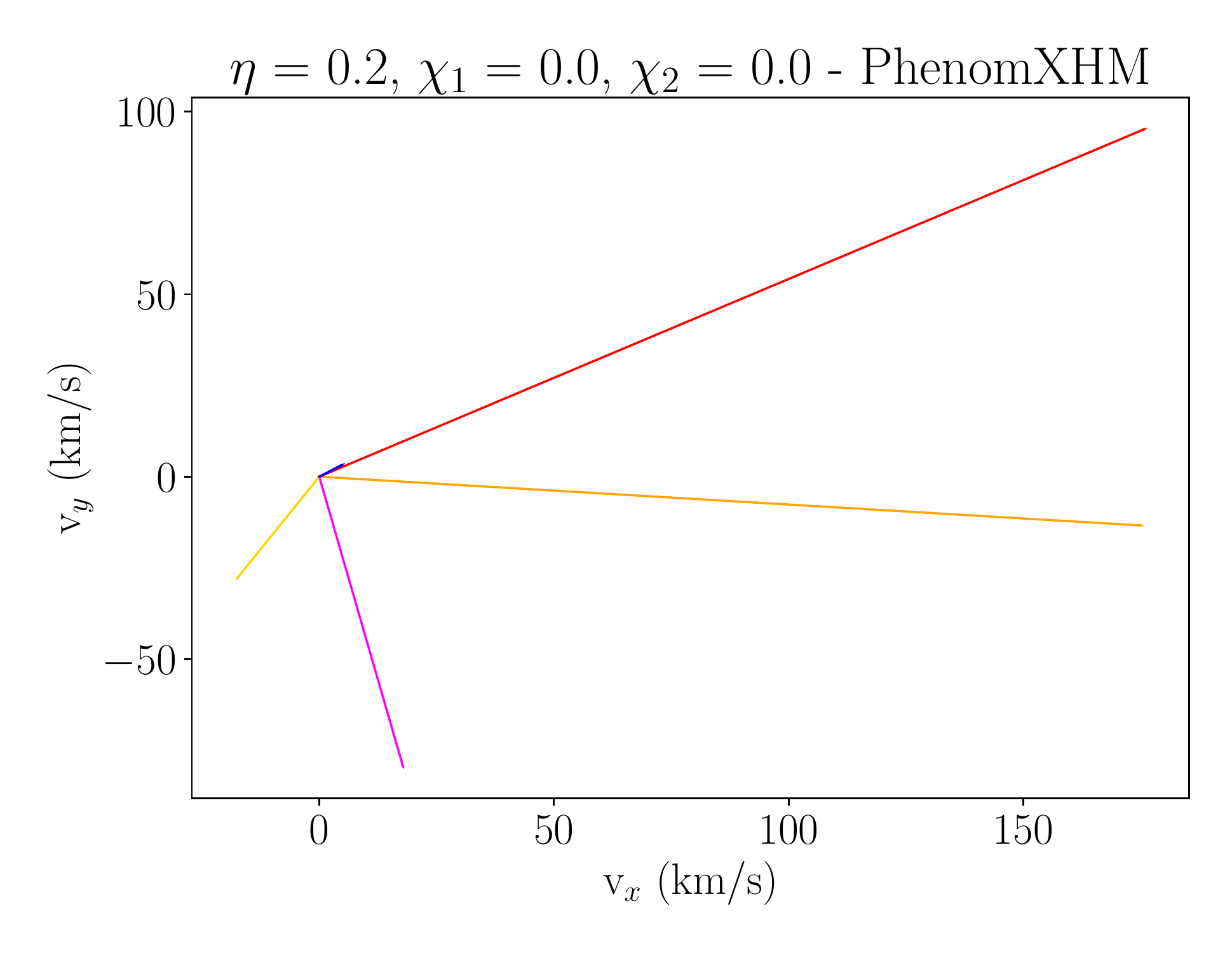}  			
		\end{subfigure}
		\begin{subfigure}{.3\textwidth}
			\centering
			\includegraphics[scale=0.26]{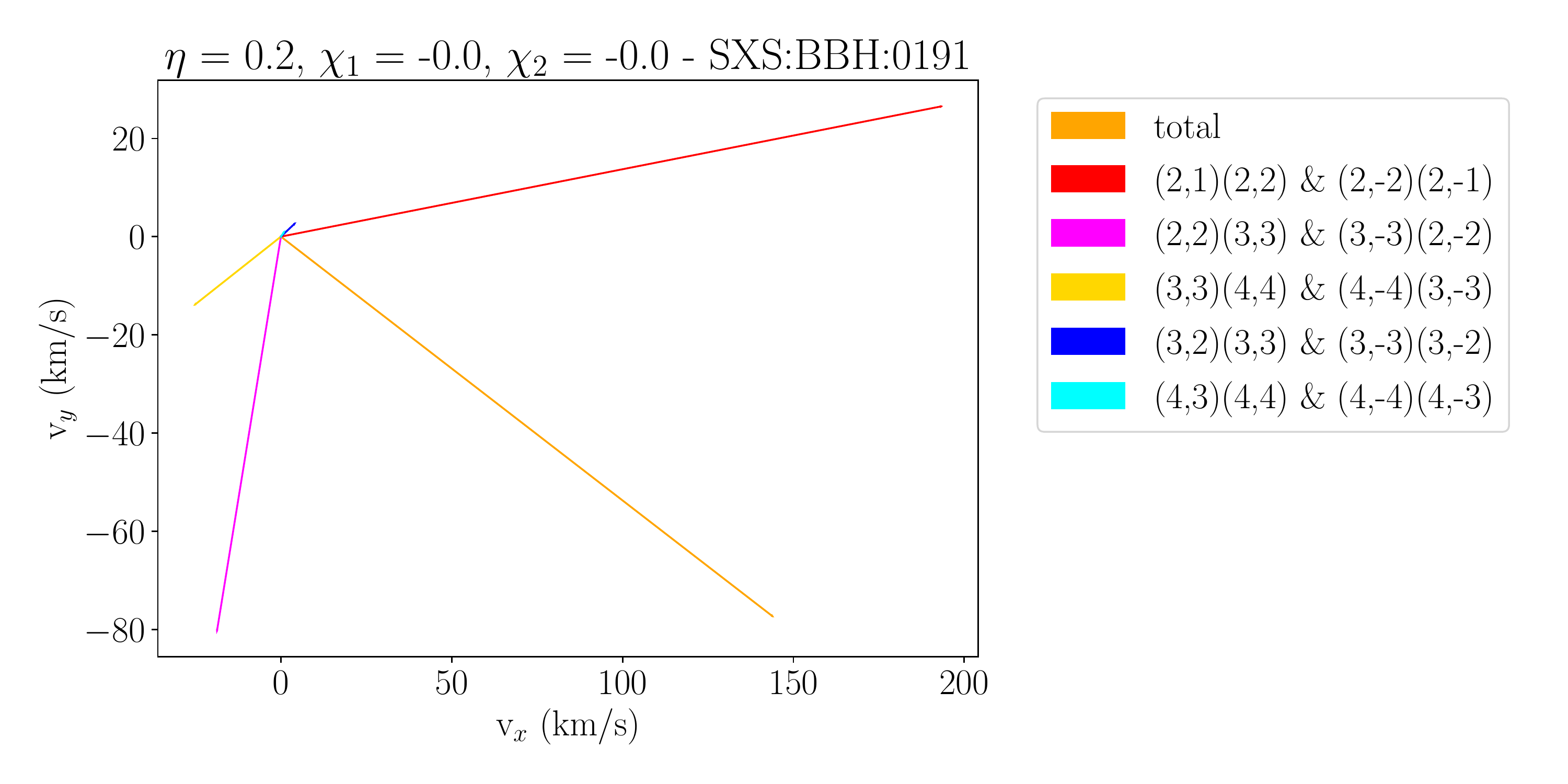}  			
		\end{subfigure}
		\caption{Comparison of the pair-contributions predicted by the two phenomenological higher-mode models and an SXS waveform, for $\{ \eta=0.2, \chi_1=0, \chi_2=0 \}$.}
		\label{pair_contributions}
	\end{figure}
	As shown in Fig. \ref{pair_contributions}, the final kick value is influenced by the magnitude and orientation of the individual contributions. Looking at the pair-contributions is thus relevant to take a step further into finding which modes show inaccuracies. The kick is highly sensitive to small time and phase shifts. Hence, minor inaccuracies can lead to significant differences in the kick predictions.
	
	\section{Kick measurements in O3a}
	The recoil velocity is an observable that can, in principle, be inferred from GW events. We investigate whether the kick velocity can be measured from the O3a events. We employ the NR surrogate fit \cite{surfinBH} on posterior samples from the GWTC-2, to obtain a posterior distribution of the kick for each event. In this case, the kick is inferred from the intrinsic properties of the binary, not from the waveform asymmetries encoded in the signal. We observe that so far, not much information can be extracted from the observed events. As an example, we show the results obtained for the event GW190412, the first GW event with highly asymmetric masses. 
	\begin{table}[h!]
		\begin{center}
			\begin{tabular}{ c c  } 
				\hline
				\hline
				\textit{waveform model} & $v$ (km/s) \\
				\hline
				IMRPhenomD & 122 $\pm$ 30\\ 
				
				IMRPhenomHM & 123 $\pm$ 27\\ 
				
				NRHybSur3dq8 & 106 $\pm$ 16 \\ 
				
				SEOBNRv4HM\_ROM & 108 $\pm$ 17\\
				
				SEOBNRv4\_ROM & 115 $\pm$ 24\\
				\hline
				IMRPhenomPv3 & 431 $\pm$ 289 \\
				
				IMRPhenomPv3HM & 382 $\pm$ 228 \\
				
				SEOBNRv4P & 316 $\pm$ 173 \\
				
				SEOBNRv4PHM & 287 $\pm$ 146\\
				\hline
				\hline
			\end{tabular}
		\end{center}
		\caption{\label{190412_values} Kick velocity values inferred by the specified waveform models for GW190412.}
	\end{table}

The first clear distinction one observes is the significant difference between considering a precessing or non-precessing waveform model for the analysis. The main reason is, non-precessing systems always have small kicks ($v_{max} \sim 300$ km/s), while precessing systems can have a broader range of velocity values. This means, one should always consider a precessing model to infer the kick in order not to restrict the range of possible values. Besides, mode asymmetries play an important role in the generation of the recoil for precessing systems. However, existing effective-one-body and phenomenological GW models do not include the mode asymmetries responsible for the highest kicks. Apart from precession effects, we observe that those models which include higher harmonics can infer the recoil more accurately. This is because waveform asymmetries are described in terms of the dominant and higher multipoles. In other words, the final velocity is built up by the sum of mode contributions. Hence, higher-mode models can give a better description of the kick. 

\section{Summary}
The kick is a sensitive tool to measure waveform accuracy. In particular, we find that higher-mode models used in current GW studies are not consistent in their kick predictions. Based on the kick harmonic contributions, we observe that waveform inaccuracies could be tuned by applying small phase and time shifts to particular modes. Our work is the starting point to improve the accuracy of higher-mode models, which in turn, can lead to meaningful kick measurements in the future.

				\end{document}